\documentclass[aps,prb,preprint,superscriptaddress]{revtex4-1}
\usepackage{color}
\usepackage{graphicx}
\usepackage{lineno}
\usepackage[normalem]{ulem}
\usepackage[colorinlistoftodos]{todonotes}

\begin{document}
\title{Evolution of spin excitations from bulk to monolayer FeSe}

\author{Jonathan Pelliciari}
\email[email: ]{pelliciari@bnl.gov}
\affiliation{Department of Physics, Massachusetts Institute of Technology, Cambridge, MA 02139, USA}
\affiliation{NSLS-II, Brookhaven National Laboratory, Upton, NY 11973, USA}

\author{Seher Karakuzu}
\affiliation{Center for Nanophase Materials Sciences, Oak Ridge National Laboratory, Oak Ridge, Tennessee 37831-6164, USA}

\author{Qi Song}
\affiliation{State Key laboratory of Surface Physics and Department of Physics, Fudan University, Shanghai, 200433, China}

\author{Riccardo Arpaia}
\affiliation{Dipartimento di Fisica, Politecnico di Milano, I-20133 Milano, Italy}
\affiliation{Quantum Device Physics Laboratory, Department of Microtechnology and Nanoscience, Chalmers University of Technology, SE-41296 Göteborg, Sweden}

\author{Abhishek Nag}
\affiliation{Diamond Light Source, Harwell Campus, Didcot OX11 0DE, United Kingdom}

\author{Matteo Rossi}
\affiliation{Dipartimento di Fisica, Politecnico di Milano, I-20133 Milano, Italy}

\author{Jiemin Li}
\affiliation{Diamond Light Source, Harwell Campus, Didcot OX11 0DE, United Kingdom}

\author{Tianlun Yu}
\affiliation{State Key laboratory of Surface Physics and Department of Physics, Fudan University, Shanghai, 200433, China}

\author{Xiaoyang Chen}
\affiliation{State Key laboratory of Surface Physics and Department of Physics, Fudan University, Shanghai, 200433, China}

\author{Rui Peng}
\affiliation{State Key laboratory of Surface Physics and Department of Physics, Fudan University, Shanghai, 200433, China}

\author{Mirian García-Fernández}
\affiliation{Diamond Light Source, Harwell Campus, Didcot OX11 0DE, United Kingdom}

\author{Andrew C. Walters}
\affiliation{Diamond Light Source, Harwell Campus, Didcot OX11 0DE, United Kingdom}

\author{Qisi Wang}
\affiliation{State Key laboratory of Surface Physics and Department of Physics, Fudan University, Shanghai, 200433, China}

\author{Jun Zhao}
\affiliation{State Key laboratory of Surface Physics and Department of Physics, Fudan University, Shanghai, 200433, China}

\author{Giacomo Ghiringhelli}
\affiliation{Dipartimento di Fisica, Politecnico di Milano, I-20133 Milano, Italy}
\affiliation{CNR-SPIN, Dipartimento di Fisica, Politecnico di Milano, I-20133 Milano, Italy}

\author{Donglai Feng}
\affiliation{State Key laboratory of Surface Physics and Department of Physics, Fudan University, Shanghai, 200433, China}

\author{Thomas A. Maier}
\affiliation{Computational Sciences and Engineering Division, Oak Ridge National Laboratory, Oak Ridge, Tennessee 37831-6164, USA}
\affiliation{Center for Nanophase Materials Sciences, Oak Ridge National Laboratory, Oak Ridge, Tennessee 37831-6164, USA}

\author{Ke-Jin Zhou}
\affiliation{Diamond Light Source, Harwell Campus, Didcot OX11 0DE, United Kingdom}

\author{Steven Johnston}
\affiliation{Department of Physics and Astronomy, The University of Tennessee, Knoxville, TN 37996, USA}

\author{Riccardo Comin}
\email[email: ]{rcomin@mit.edu}
\affiliation{Department of Physics, Massachusetts Institute of Technology, Cambridge, MA 02139, USA}


\begin{abstract}
\textbf{The discovery of enhanced superconductivity (SC) in FeSe films grown on SrTiO$_3$ (FeSe/STO) has revitalized the field of Fe-based superconductors~\cite{huang_monolayer_2017,lee_routes_2018,wang_interface-induced_2012,peng_tuning_2014,ding_high-temperature_2016}. In the ultrathin limit, the superconducting transition temperature T$_c$ is increased by almost an order of magnitude, raising new questions on the pairing mechanism. As in other unconventional superconductors, antiferromagnetic spin fluctuations have been proposed as a candidate to mediate SC in this system~\cite{chubukov_pairing_2012,chubukov_itinerant_2015,chubukov_iron-based_2015,lee_routes_2018,linscheid_high_2016,mishra_s_2016,shigekawa_dichotomy_2019}. Thus, it is essential to study the evolution of the spin dynamics of FeSe in the ultrathin limit to elucidate their relationship with superconductivity. Here, we investigate and compare the spin excitations in bulk and monolayer FeSe grown on STO using high-resolution resonant inelastic x-ray scattering (RIXS) and quantum Monte Carlo (QMC) calculations. Despite the absence of long-range magnetic order, bulk FeSe displays dispersive magnetic excitations reminiscent of other Fe-pnictides. Conversely, the spin excitations in FeSe/STO are gapped, dispersionless, and significantly hardened relative to the bulk counterpart. By comparing our RIXS results with simulations of a bilayer Hubbard model, we connect the evolution of the spin excitations to the Fermiology of the two systems. The present study reveals a remarkable reconfiguration of spin excitations in FeSe/STO, which is essential to understand the role of spin fluctuations in the pairing mechanism.}
\end{abstract}
\maketitle

\begin{figure}
\includegraphics[width=0.75\textwidth,trim={0 0 0 0},clip]{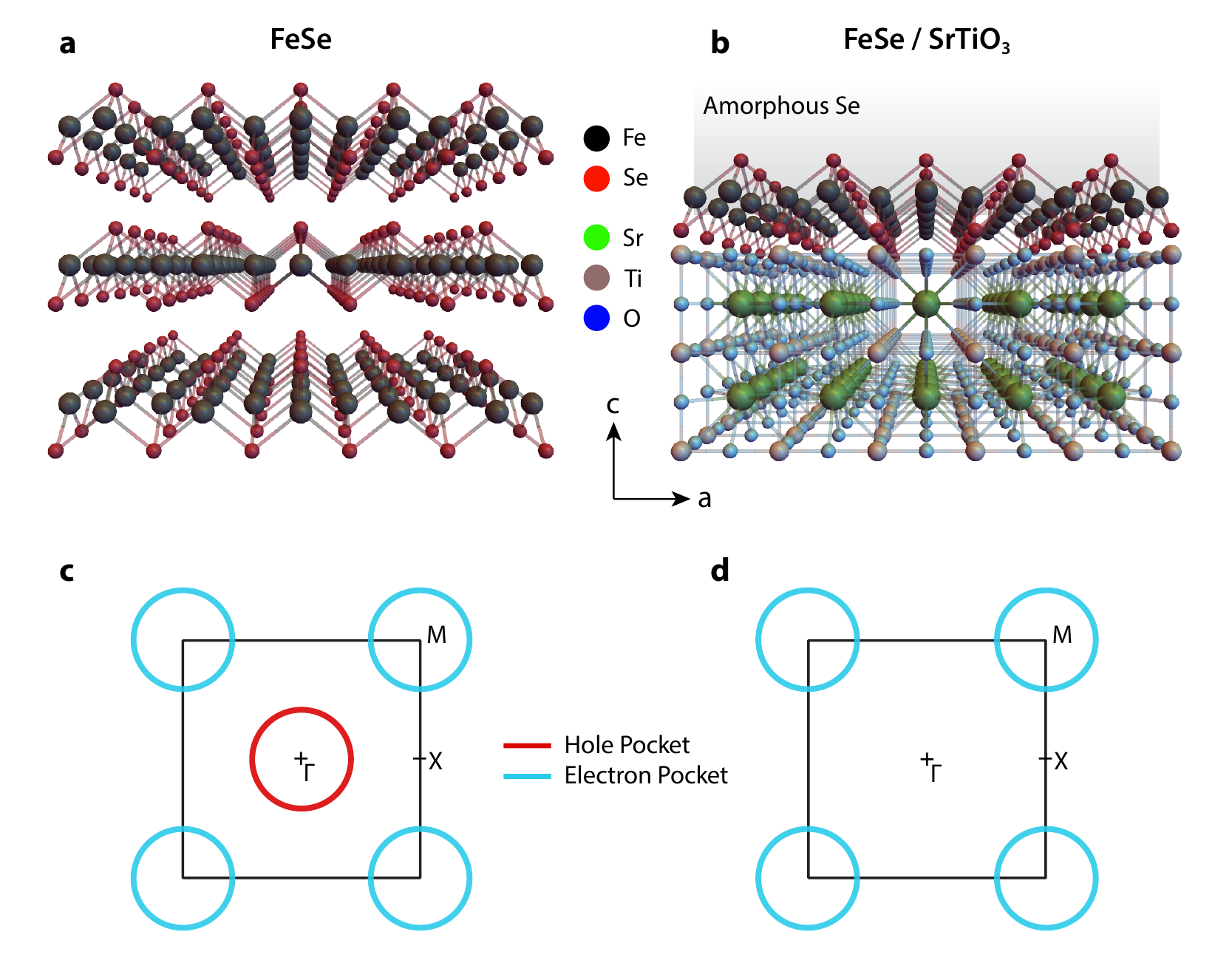}
\caption{\label{fig:fig1} \textbf{Structure and Fermi surface of FeSe bulk and FeSe/STO}. \textbf{a}: Structure of FeSe bulk. \textbf{b}: Structure of FeSe/STO monolayer with Se capping. \textbf{c,d}: Schematic Fermi surface of FeSe bulk (\textbf{c}) and FeSe/STO monolayer (\textbf{d}). The electron pocket of bulk FeSe has been drawn circular and not elliptical for simplicity and for correspondence with the theoretical model adopted here.}
\end{figure}

Iron selenide (FeSe) occupies a somewhat unique place among Fe-based superconductors. It has the simplest structure, consisting of a square Fe lattice with Se ions situated above and below it, as depicted in Fig.~\ref{fig:fig1}\textbf{a}. It is superconducting with T$_c\sim 8$\,K and has a structural transition T$_s\sim 90$\,K~\cite{dai_antiferromagnetic_2015,chen_anisotropic_2019,wang_magnetic_2016}. The Fermi surface of bulk FeSe is composed of cylindrical hole pockets at the $\Gamma$ point and elliptical electron pockets at the $M$ point (see Fig.~\ref{fig:fig1}\textbf{c}; here, a Brillouin zone with two Fe sites per unit cell  has been adopted). 
The Fermi surface of FeSe/STO, on the other hand, is composed solely of circular electron pockets at the $M$ point~\cite{zhang_superconducting_2016,huang_monolayer_2017,lee_routes_2018}, while the hole pockets at the $\Gamma$ point are pushed below the Fermi level (Fig.~\ref{fig:fig1}\textbf{d}). These observations are consistent with an electron doping of $\sim 0.1 / \mathrm{Fe}$, as extracted from the Luttinger count~\cite{zhang_superconducting_2016,huang_monolayer_2017,lee_routes_2018}, suggesting that STO acts as an electron donor for monolayer FeSe. 

Simultaneous N{\'e}el- and stripe-like fluctuations have been observed in bulk FeSe at ${\bf q} = (1,0)$ and $(1,1)$ (reciprocal lattice units, r.l.u.), despite the lack of long-range antiferromagnetic order. These observations signal the presence of significant magnetic frustration that ultimately precludes any long-range order~\cite{wang_magnetic_2016}.
From an experimental perspective, the investigation of spin excitations in FeSe/STO is complicated by the limited volume contributing to the magnetic scattering signal. Inelastic neutron scattering (INS) is currently unable to probe single atomic layers, and other light scattering techniques, such as Raman and optical spectroscopy, cannot disentangle the signals from the substrate, the FeSe layer, and the interface between the two. On this front, recent advances in Resonant Inelastic X-ray Scattering (RIXS) have allowed the detection of spin excitations in Fe-based superconductors, producing complementary information to INS~\cite{zhou_persistent_2013,pelliciari_presence_2016,pelliciari_intralayer_2016,pelliciari_local_2017,rahn_paramagnon_2019,pelliciari_reciprocity_2019,garcia_anisotropic_2019,ament_resonant_2011,dean_insights_2015}. The signal enhancement and sensitivity to electronic excitations that is afforded by resonant photoexcitation render RIXS a prime technique for investigating ultrathin materials. Additionally, the elemental selectivity of RIXS enables one to isolate the signal from specific atoms and disentangle the contributions from the film and the substrate. These aspects make RIXS an ideal technique for studying magnetic excitations in FeSe/STO. 

Here, we combine high-energy-resolution RIXS measurements and quantum Monte Carlo calculations within the dynamical cluster approximation (DCA) to elucidate the spin dynamics of bulk FeSe and FeSe/STO films down to the single unit cell limit. We find that the magnetic excitations in FeSe/STO are gapped and dispersionless in momentum space, and harden significantly relative to other Fe-based superconductors. These observations are in stark contrast with the spin excitations of bulk FeSe, which exhibit an acoustic-like dispersion toward the zone center, similarly to other antiferromagnetic systems \cite{dai_antiferromagnetic_2015}. The evolution of the spin excitations is captured by DCA calculations of a bilayer Hubbard model~\cite{maier_pair_2011}, which accounts for the transition from a two-band system into an incipient band system (see Methods). Correspondingly, we establish that the reconfiguration of the spin excitations from bulk to monolayer FeSe originates from the Lifshitz transition of the Fermi surface and accompanying loss of the hole pocket at the $\Gamma$ point. This transition quenches particle-hole scattering processes, flattens and gaps out their dispersion, and increases their energy bandwidth, in agreement with the experimental observations. 

\begin{figure}
\includegraphics[width=0.9\textwidth,trim={0 0 0 0},clip]{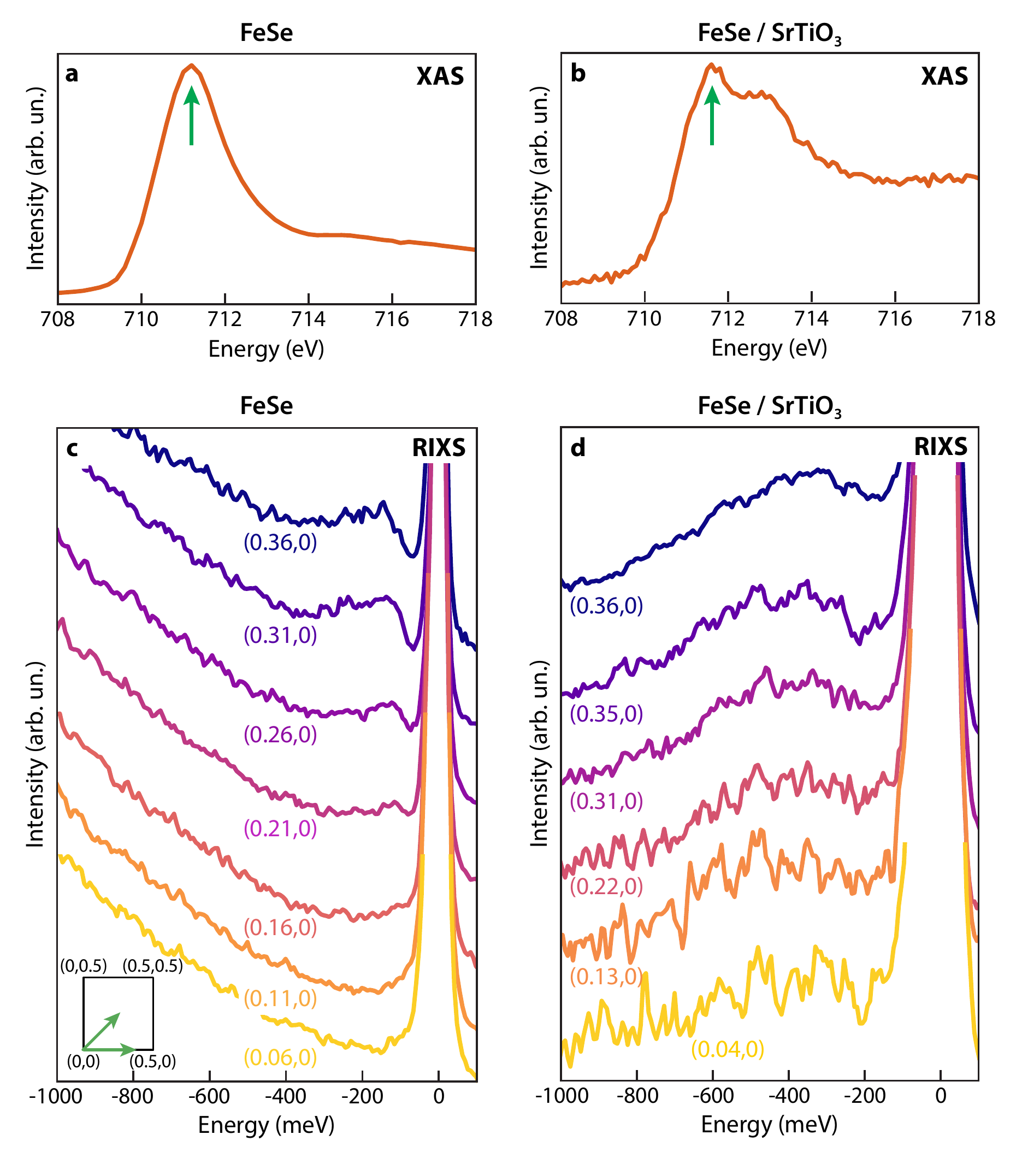}
\caption{\label{fig:fig2} \textbf{XAS and RIXS spectra for FeSe bulk and Fe/STO.} \textbf{a,b}: Fe $L_3$-edge X-ray absorption spectra for FeSe bulk (\textbf{a}) and FeSe/STO (\textbf{b}), measured via total electron yield. The arrows mark the incident energy for the RIXS data displayed in \textbf{c} and \textbf{d}. \textbf{c,d}: High-energy resolution RIXS spectra of FeSe bulk (\textbf{c}) and FeSe/STO (\textbf{d}) at different momentum points along the high-symmetry direction $(0,0) \rightarrow (H,0)$ [RIXS spectra along the $(0,0) \rightarrow (H,H)$ direction are reported in the Supplementary Information].}
\end{figure}

Figures~\ref{fig:fig2}\textbf{a} and~\ref{fig:fig2}\textbf{b} summarize the Fe $L$-edge X-ray Absorption Spectroscopy (XAS) data for bulk FeSe (FeSe hereafter) and monolayer FeSe (FeSe/STO hereafter), respectively. The XAS of FeSe resembles the spectra previously obtained from cleaved Fe pnictides crystals with Fe in a $2+$ oxidation state and embedded in a metal environment \cite{pelliciari_presence_2016,pelliciari_reciprocity_2019,zhou_persistent_2013,pelliciari_local_2017,garcia_anisotropic_2019}. The XAS of FeSe/STO has an additional peak at higher energy, which could originate from new interfacial valence states induced by hybridization with orbitals of the STO substrate. The arrows in Figs.~\ref{fig:fig2}\textbf{a},\textbf{b} specify the incident photon energies at which RIXS spectra were collected. 

Figures~\ref{fig:fig2}\textbf{c} and \ref{fig:fig2}\textbf{d} show the corresponding high-resolution and high-statistics RIXS data on FeSe and monolayer FeSe/STO, respectively. In the bulk case, we detect a dispersive excitation at an energy of $\sim140$ meV at ${\bf q} = (0.36,0)$ r.l.u., which gradually decreases in energy toward the zone center until it merges into the elastic line. This mode is reminiscent of what observed in INS experiments~\cite{wang_magnetic_2016} and can be ascribed to spin excitations as previously shown in Ref.~[\onlinecite{rahn_paramagnon_2019}]. A word of caution should be given, however, as FeSe lacks long-range antiferromagnetism and instead exhibits N{\'e}el- and stripe-type fluctuations \cite{wang_magnetic_2016}. As such, a direct comparison between the excitations measured by INS and RIXS is not straightforward since the $\Gamma$ point is not equivalent to $M$ or $X$ in the absence of Brillouin zone folding. Nevertheless, the excitations of FeSe closely resemble those observed in BaFe$_2$As$_2$ \cite{rahn_paramagnon_2019}, suggesting that spin fluctuations are of similar nature in these two compounds  in proximity of the $\Gamma$ point and across the portion of Brillouin zone accessible to RIXS.

\begin{figure}
\centering
\includegraphics[width=0.75\textwidth,trim={0 0 0 0},clip]{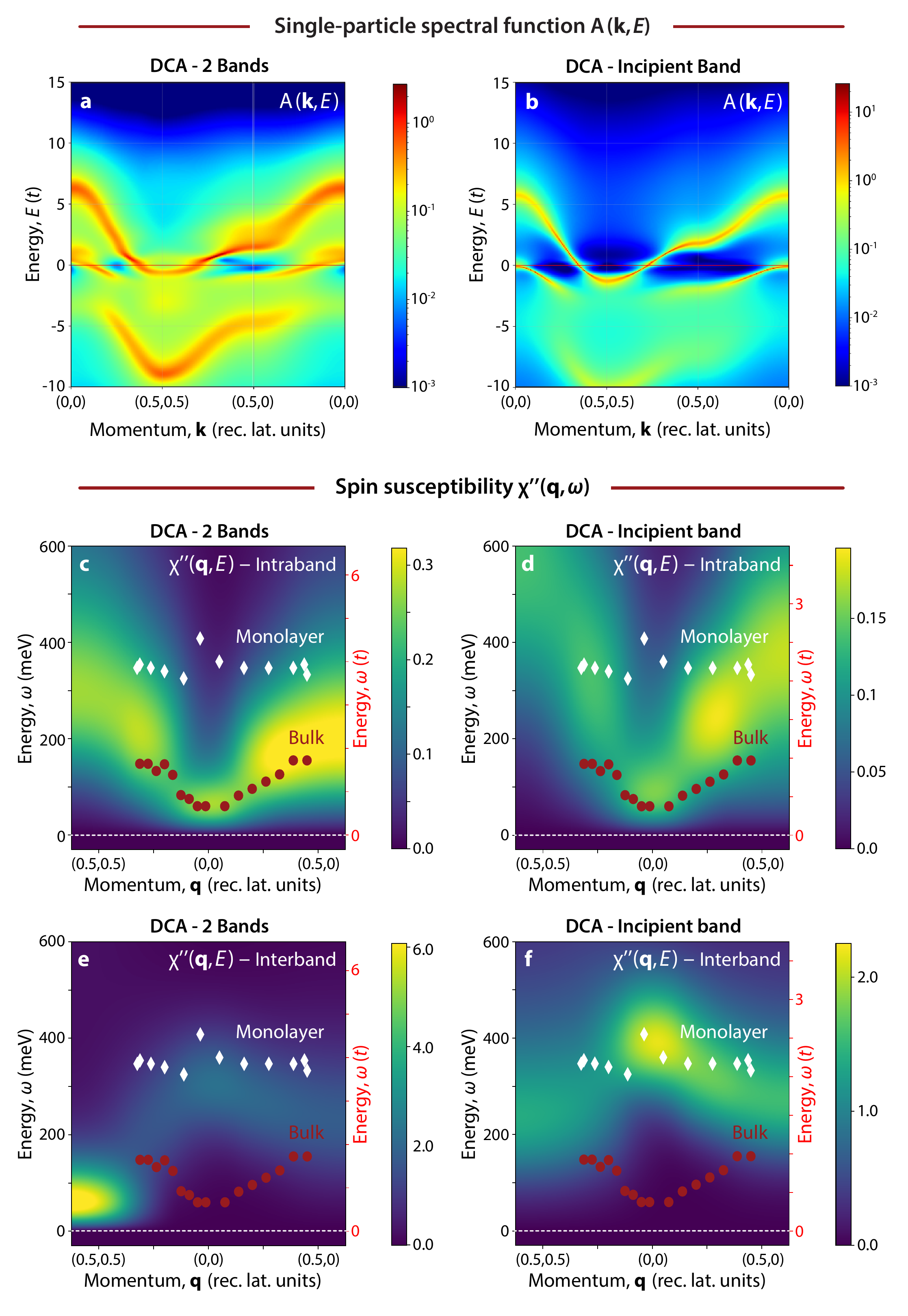}
\caption{\label{fig:fig3} \textbf{Single-particle spectral function and dynamical spin susceptibility from DCA calculations.} \textbf{a,b}: Dynamical cluster approximation (DCA) calculations and spectral function $A({\bf k},E)$ for the two-band Hubbard model (\textbf{a}) and the incipient band Hubbard model (\textbf{b}). \textbf{c-f}: DCA calculations of the imaginary part of the spin susceptibility $\chi^{\prime\prime}({\bf q},\omega)$ for the two-band Hubbard model (\textbf{c}: intraband  Q$_z = 0$; \textbf{e}: interband Q$_z = \pi$) and the incipient band Hubbard model (\textbf{d}: intraband Q$_z = 0$; \textbf{f}: interband Q$_z = \pi$). Red circles (white diamonds) indicate the energy position of the peak detected by RIXS in bulk (monolayer) FeSe. The uncertainties associated with peak fitting are smaller than the markers.}
\end{figure}

We observe significant differences in the RIXS spectra collected on the FeSe monolayer. At zero energy loss we detect a strong elastic signal that likely reflects the overall diffuse scattering from the capping layer, the FeSe film, and the STO substrate. Despite this strong elastic background, we are able to identify inelastic peaks owing to the high energy resolution of the instrument ($\sim40$ meV). In particular, we observe a broad peak located at $\sim320$ meV at ${\bf q} = (0.36,0)$ r.l.u., whose energy linewidth is significantly greater than the excitations detected in the bulk case. This peak is largely asymmetric -- similar to bulk FeSe -- but its tail extends to energies as high as $1$ eV, much higher than the bulk counterpart. Furthermore, this mode barely disperses as a function of momentum and has an energy of $\sim320-400$ meV along the $(H,0)$ and $(H,H)$ directions, as reported in Fig.~\ref{fig:fig2}\textbf{d} and~\ref{fig:fig3}. Thanks to resonant photoexcitation at the Fe-$L$ edge, we can identify the FeSe layer as the host of this excitation. This interpretation is further supported by the dependence of the RIXS signal on the incident photon energy across the resonance (see Supp. Inf.). The ability to make this assignment is essential to disentangle excitations originating from the film, the substrate or the interface.

The evolution of the spin excitations from FeSe bulk to monolayer is significant and cannot be compared nor ascribed to any doping effects previously observed in related materials. For example, the spin excitations of BaFe$_2$As$_2$ evolve differently depending on the doping type: in the case of hole doping (K-), the spin excitations gradually soften upon doping \cite{wang_doping_2013,zhou_persistent_2013}, while electron doping (Co/Ni-) leaves the high-energy spin excitations more or less unaffected \cite{luo_electron_2012,luo_electron_2013,wang_doping_2013}. The case of isovalent doping (P-), where the spin excitations harden gradually \cite{hu_spin_2016,pelliciari_reciprocity_2019}, is also interesting. Nonetheless, the doping-induced changes observed in these systems are minor compared to the effect observed here. The hardening of spin excitations measured in P-doped BaFe$_2$As$_2$ -- so far the largest reported in the literature -- is much smaller ($40$ meV) than what we observe in FeSe. Most importantly, a clear dispersion is found in these compounds at all doping levels, contrary to the flat momentum dependence in the FeSe monolayer. 

The principal difference between FeSe and FeSe/STO is in their band structure and Fermi surface topology. To explore the impact of these differences on the spin excitations, we calculated the single-particle spectral function $A({\bf k},E)$ and dynamical spin susceptibility $\chi^{\prime\prime}({\bf q},\omega)$ of the bilayer Hubbard model using the dynamical cluster approximation (DCA) and a nonperturbative quantum Monte Carlo solver (see Methods). The bilayer Hubbard model is the simplest model with an electronic structure similar to the Fe-based superconductors that can be studied with QMC while maintaining a manageable sign problem. By varying the value of the nearest-neighbour interlayer hopping  $t_\perp$, the electronic structure of the model can be tuned from a system with both hole- and electron-like bands crossing the Fermi level (Fig.~\ref{fig:fig3}\textbf{a}) to one with a single electron-like band crossing the Fermi level and an incipient hole band (Fig.~\ref{fig:fig3}\textbf{b}). The model can, therefore, capture the qualitative features of the band structure of bulk and monolayer FeSe. In Fig.~\ref{fig:fig3}\textbf{a}, we report the spectral function for the two-band model, where we observe a hole-like band crossing the Fermi level close to the $\Gamma$ point and an electron-like band intersecting the Fermi level in proximity of the $M$ point. This band structure leads to a double pocket Fermi surface as sketched in Fig.~\ref{fig:fig1}\textbf{c}. In the case of the incipient band model, shown in Fig.~\ref{fig:fig3}\textbf{b}, the hole band at the $\Gamma$ point is pushed to lower energies, moving below the Fermi level and removing the hole pocket at the $\Gamma$ point. The resulting Fermi surface is composed only of a circular electron pocket at the $M$ point, as sketched in Fig.~\ref{fig:fig1}\textbf{d}. 

Figures~\ref{fig:fig3}\textbf{c-f} display the calculated imaginary part of the spin susceptibility $\chi^{\prime\prime}({\bf q},\omega)$ spectra for two values of $t_\perp$, corresponding to bulk and FeSe/STO. In our model, two components of $\chi^{\prime\prime}({\bf q},\omega)$ are extracted with intra- ($q_\perp = 0$) and interband ($q_\perp = \pi$) character, which can be isolated from one another by choosing the appropriate value of $q_\perp$. Figures~\ref{fig:fig3}\textbf{c-f} report the intra- and interband channels in the middle and bottom rows, respectively. In the case of the two band model with two ambipolar Fermi pockets, we obtain a strongly dispersing $\chi^{\prime\prime}({\bf q},\omega)$ (see Fig.~\ref{fig:fig3}\textbf{c,e}), whose main two components  -- arising from intraband and interband scattering -- are dispersing out-of-phase in momentum space. Specifically, the intraband component has a minimum at the $\Gamma$ point and increases in energy towards its maximum at $(0.5,0)$ and $(0.5,0.5)$ while the interband component displays two minima at $(0.5,0)$ and $(0.5,0.5)$ and a maximum at $(0,0)$. An analysis of the spectral intensity reveals that the interband component is four to five times larger than the intraband one.

Upon increasing $t_\perp$, the hole-like band is made incipient. The interband component of the resulting $\chi^{\prime\prime}({\bf q},\omega)$ is much less dispersive and becomes gapped throughout the entire Brillouin zone, in close agreement with the experimental findings (see Fig.~\ref{fig:fig3}\textbf{f}). The out-of-phase dispersion of the intra- and interband $\chi^{\prime\prime}({\bf q},\omega)$ is also preserved for the incipient band condition. The calculation additionally captures the broadening of the peaks in the incipient band case compared to the two-band model.

Figures~\ref{fig:fig3}\textbf{c-f} additionally summarize our results by comparing the calculated inter- and intra-band $\chi^{\prime\prime}({\bf q},\omega)$ as a false color image, with experimental peak positions overlaid. Here, the results are shown for both bulk (white circles) and monolayer (white diamonds) FeSe (a more detailed description of the extraction of the experimental data points is given in the Supp. Inf.). We have assumed $t=90$ ($160$) meV for the bulk (incipient) case when converting the DCA energy scale to physical units, which produces the best agreement with the experimental data. The use of different factors for the two cases is supported by recent DMFT+LDA calculations, which indicate that bulk FeSe is more correlated than the FeSe/STO~\cite{mandal_how_2017} (this conclusion is also consistent with our observation of much sharper spectral functions in the incipient band case, see Fig.~\ref{fig:fig3}\textbf{a,b}). Based on this, it is natural to adopt a larger \textit{t} for the monolayer case while holding the value of \textit{U} fixed. 

The experimental dispersions in bulk FeSe appear to be in better agreement with the intraband $\chi^{\prime\prime}({\bf q},\omega)$ (Fig.~\ref{fig:fig3}\textbf{c,d}) rather than the interband component (Fig.~\ref{fig:fig3}\textbf{e,f}). The intensity of the interband $\chi^{\prime\prime}({\bf q},\omega)$ is higher than the intraband $\chi^{\prime\prime}({\bf q},\omega)$ and one might expect that the RIXS signal scales proportionally. However, matrix elements of the RIXS cross section have not been included in the model, which makes a qualitative comparison the only viable option (including Fe-L edge matrix elements would require a momentum-resolved full multi-orbital Fe calculation, which is currently not possible  due to the severe Fermion sign problem induced by Hund's coupling). In any case, from a phenomenological perspective, the agreement with the intraband $\chi^{\prime\prime}({\bf q},\omega)$ is good and future calculations including orbital orientation and polarization effects could offer a more quantitative description of the RIXS cross section. In Figs.~\ref{fig:fig3}\textbf{d,f}, we report the calculations obtained for the incipient band model (tailored for FeSe/STO), where the agreement between theory and experiments is better for the interband $\chi^{\prime\prime}({\bf q},\omega)$. In this case, the interband $\chi^{\prime\prime}({\bf q},\omega)$ is flattened by the lack of the hole pocket and a hardening of the dispersion is reproduced by the theory. 
These changes are a direct consequence of the fact that intraband scattering is strongly suppressed at low-energies once the hole pocket is shifted below the Fermi level. This hardening and flattening of the electronic excitations is clearly observed in the experimental data for FeSe/STO as corroborated by the diamonds overlaid with the color plot. The interband $\chi^{\prime\prime}({\bf q},\omega)$ also has the largest intensity compared to the intraband $\chi^{\prime\prime}({\bf q},\omega)$, and is, therefore, expected to dominate the RIXS signal when neglecting cross section effects. 

Our findings have implications for the enhancement of SC in FeSe/STO. In Eliashberg- and fluctuation exchange-type models (FLEX), $\chi^{\prime\prime}({\bf q},\omega)$ enters directly into the equation to calculate T$_c$ \cite{maier_pair_2011,linscheid_high_2016}. The significant evolution in $\chi^{\prime\prime}({\bf q},\omega)$ revealed by RIXS data suggests a sizable change in this section of the equation, highlighting the importance of spin excitations for a complete explanation and description of SC in FeSe/STO. Moreover, any quantitative model for the spin fluctuation contribution to pairing must also account for the observed evolution of the spin dynamics. As such, the evolution of the spin dynamics from FeSe to FeSe/STO represents an essential clue to a magnetic-like pairing scenario, which was previously proposed for other Fe pnictides \cite{chubukov_pairing_2012,dai_antiferromagnetic_2015,linscheid_high_2016}. The present results do not, however, rule out additional interactions such as phonons or doping from the substrate, which can contribute to the enhancement of T$_c$ \cite{lee_interfacial_2014,zhang_ubiquitous_2017}. 

In summary, we report a combined experimental and theoretical investigation of the spin dynamics in bulk FeSe and single-unit-cell FeSe/STO, uncovering a dramatic evolution of magnetic excitations from the bulk to the monolayer limit. In bulk FeSe, we observed dispersive spin excitations that are reminiscent of other Fe-based superconductors. These modes become significantly more energetic and less dispersive in the ultrathin limit of the FeSe/STO film. Quantum  Monte Carlo calculations of the bilayer Hubbard model reveal that this reconfiguration of spin dynamics is a direct consequence of the suppression of the interband scattering once the hole pocket is removed from the Fermi level. These findings suggest a fundamental link between the Fermiology of FeSe superconductors and their spin dynamics up to a very high energy scale. The direct experimental insights of the present RIXS study underscore the role of spin excitations for unconventional SC in FeSe, and provide an empirical benchmark for theoretical models of SC in FeSe/STO. 

\section{Methods}
\textit{Sample preparation}

\textit{Monolayer FeSe on STO} --- Monolayer of FeSe was grown on Nb-doped ($0.5\%$ wt.) (001)-oriented SrTiO$_3$ substrate. The substrate was etched following the method described in Ref.~[\onlinecite{tan_interface-induced_2013}]. In the growth chamber, which has a base pressure of $6\times 10^{-10}$ mbar, the substrate was heated to $800^\circ$C for $45$ minutes with Se flux. Single-layer FeSe films were grown at $\sim500^\circ$C by coevaporation of Se and Fe with a flux ratio of $20:1$. After growth, the films were annealed at $550^\circ$C in vacuum for 2 hours. The FeSe/STO was characterized by ARPES and the superconducting gap was determined to be $\sim 13.4$ meV or T$_c\sim 60-65$ K.
A $\sim25$ nm thick layer of amorphous Se was added for protection at room temperature.

\textit{FeSe bulk} --- Bulk FeSe single crystals were grown under a permanent gradient of temperature ($\sim400-330^\circ$C) in the KCl-AlCl$_3$ flux, as reported in Ref.~[\onlinecite{wang_magnetic_2016}]. The T$_c$ of the bulk FeSe sample is $~\sim8$ K.

\textit{High energy resolution RIXS measurements on FeSe bulk and FeSe/STO} ---  High-resolution RIXS experiments were performed at the I21-RIXS beamline at Diamond Light Source, United Kingdom. FeSe bulk was cleaved in vacuum. All samples were aligned with the surface normal (001) lying in the scattering plane. X-ray absorption was measured using the total electron yield (TEY) method by recording the drain current from the samples. For RIXS measurements, $\pi$ polarized light was used. The combined energy resolution was about 40 meV (FWHM) at the Fe $L_3$ edge ($\sim710.5$ eV). To enhance the RIXS throughput, a parabolic mirror has been installed in the main vacuum chamber. The RIXS spectrometer was positioned at a fixed scattering angle of 154 degrees resulting in a maximal total momentum transfer value $\textbf{Q}$ of $\sim 0.7~\mathrm{\AA}^{-1}$. The projection of the momentum transfer, \textbf{q}, in the \textit{ab} plane was obtained by varying the incident angle on the sample. We use the 2 Fe unit cell convention with $a = b = 3.76$~\AA~and $c = 5.4$~\AA~for the reciprocal space mapping. The momentum transfer ${\bf Q}$ is defined in reciprocal lattice units (r.l.u.) as ${\bf Q} = {\bf H}a^* + {\bf K}b^* + {\bf L}c^*$ where $a^* = 2\pi/a$, $b^* = 2\pi/b$, and $c^* = 2\pi/c$. All measurements were performed at 20 K under a vacuum pressure of about $5\times 10^{-10}$ mbar.

Spectra for the FeSe have been acquired in $\sim30$ minutes whereas spectra for the FeSe/STO required 3 hours or more for every momentum point.

\textit{Calculations} --- We modeled the spin excitation spectrum of bulk and monolayer FeSe using a two-orbital Hubbard model defined on a two-dimensional square lattice with $N=L^2$ unit cells, where $L$ is the linear size of the system. This model includes only the intraorbital Hubbard repulsion $U$ on each orbital, and it is identical to the one used in Ref.~ [\onlinecite{maier_pair_2011}]. (Details are also provided in the Supp. Inf. for completeness). Due to the orbital symmetry of the Hamiltonian, and the restriction to only a local intra-orbital Hubbard interaction, one can regard this model as a bilayer Hubbard model with layers $\alpha = 1,2$.~\cite{maier_pair_2011} The kinetic energy term can then be diagonalized 
and rewritten in terms of a bonding $k_z=0$ and anti-bonding $k_z=\pi$ basis. As such, momentum transfers with $q_z = 0$ and $\pi$ correspond to intra- and interband excitations, respectively. Throughout, we use $t=1$ as the unit of energy, set $U=8t$, and vary $t_{\perp}$ and the filling $n$ to control the electronic structure of the system. 

We simulated the model using the dynamical cluster approximation (DCA) method~\cite{maier_quantum_2005}, 
where the bulk lattice system is mapped onto a periodic finite-size cluster embedded in a mean-field. The effective cluster problem was solved self-consistently by means of a continuous-time auxiliary field (CTAUX) quantum Monte Carlo method~\cite{hahner_dca_2020,gull_continuous-time_2008,gull_submatrix_2011}. The real frequency dynamical correlation functions shown here were obtained from QMC data using the Maximum Entropy (MaxEnt) method~\cite{gubernatis_quantum_1991}. 

\section{Authors contribution}
J.~P., R.~A., A.~N., M.~R., J.~L., M.~G.F., G.~G, A.~C.~W.,K.~Z. performed the RIXS experiments. Q.~S., T.~Y, X.~C., R.~P., D.~F. prepared the FeSe/STO thin films. Q.~W. and J. Z. grew FeSe single crystals. S.~K., S.~J. and T.~A.~M. performed the theory calculations. J.~P.,S.~K., S.~J., T.~A.~M., and R.C. wrote the manuscript with input from all the authors. R.~C. supervised the project.

\section{Acknowledgements}
We acknowledge John Tranquada, Rafael Fernandes, Connor Occhialini, and Andrey Chubukov for enlightening discussions. We also thank Nick Brookes, Kurt Kummer, and Davide Betto for initial tests on FeSe/STO. This work was supported by the Air Force Office of Scientific Research Young Investigator Program under grant FA9550-19-1-0063. We thank Diamond Light Source for the allocation of beamtime to proposal SP18883. J.~P. acknowledges financial support by the Swiss National Science Foundation Early Postdoc Mobility Fellowship Project No. P2FRP2\_171824 and P400P2\_180744. S.~K., T.~A.~M., and S.~J. are supported by the Scientific Discovery through Advanced Computing (SciDAC) program funded by U.S. Department of Energy, Office of Science, Advanced Scientific Computing Research and Basic Energy Sciences, Division of Materials Sciences and Engineering. S. J. acknowledges additional support from the Office of Naval Research under Grant No. N00014-18-1-2675. An award of computer time was provided by the INCITE program. This research also used resources of the Oak Ridge Leadership Computing Facility, which is a DOE Office of Science User Facility supported under Contract DE-AC05-00OR22725. M.~R. and G.~G. were supported by the ERC-P-ReXS project (2016-0790) of the Fondazione CARIPLO and Regione Lombardia, in Italy.  R.A. is supported by the Swedish Research Council (VR) under the Project 2017-00382. R.C. acknowledges support from the Alfred P. Sloan Foundation.

\section{Competing Interests} 
The authors declare no competing and financial interests.

\section{Data Availability}
Data that support the findings of this study are available upon reasonable request from the corresponding authors.

 \section{Correspondence} Correspondence and requests for materials should be addressed to J. Pelliciari and R. Comin.

\clearpage

\clearpage
\end{document}